%% file: main.tex
\documentclass{article}
\usepackage{spconf,amsmath,graphicx,hyperref}
\usepackage{xcolor}
\usepackage{amssymb}

\usepackage{rotating}    
\usepackage{adjustbox}   

\usepackage[normalem]{ulem}
\usepackage{graphicx}
\usepackage{subcaption}  
\usepackage{lipsum}
\usepackage{float}
\usepackage{amssymb}

\usepackage{caption}   
\usepackage{placeins}  

\usepackage[percent]{overpic} 
\usepackage{xcolor}
\newsavebox{\myimagebox}
\input{commands}

\title{Direct Kernel Optimization: Efficient Design for Opto-Electronic Convolutional Neural Networks}
\name{%
  \begin{tabular}{c}
    Ali Almuallem$^{1}$, Harshana Weligampola$^{1}$, Abhiram Gnanasambandam$^{2}$, Wei Xu$^{1}$, \\
    Dilshan Godaliyadda$^{2}$, Hamid R. Sheikh$^{2}$, Stanley H. Chan$^{1}$, and Qi Guo$^{1}$
  \end{tabular}%
  \thanks{The work was supported by Samsung Research America.}%
}
\address{$^{1}$Elmore Family School of Electrical and Computer Engineering, Purdue University\\
$^{2}$Samsung Research America}

\begin{document}

\maketitle
\begin{abstract}

Hybrid opto-electronic neural networks combine optical front-ends with electronic back-ends to perform vision tasks, but joint end-to-end (E2E) optimization of optical and electronic components is computationally expensive due to large parameter spaces and repeated optical convolutions. We propose \textit{Direct Kernel Optimization (DKO)}, a two-stage training framework that first trains a conventional electronic CNN and then synthesizes optical kernels to replicate the first-layer convolutional filters, reducing optimization dimensionality and avoiding hefty simulated optical convolutions during optimization. We evaluate DKO in simulation on a monocular depth estimation model and show that it achieves twice the accuracy of E2E training under equal computational budgets while reducing training time. Given the substantial computational challenges of optimizing hybrid opto-electronic systems, our results position DKO as a scalable optimization approach to train and realize these systems.
\end{abstract}

\section{Introduction}

Opto-electronic neural networks work by combining optical front-ends—such as transmission masks~\cite{klotz2024minimalist}, diffractive optical elements~\cite{sitzmann2018end}, and metasurfaces~\cite{zheng_meta-optic_2022} with electronic back-ends based on conventional neural architectures to perform vision and imaging tasks. By leveraging optics to preprocess signals before electronic inference, such systems offer the potential for low-latency~\cite{lin_all-optical_2018} and energy-efficient~\cite{klotz2024minimalist} computation. However, most existing approaches rely on an end-to-end training paradigm in which both the optical components and the electronic layers are optimized jointly~\cite{sitzmann2018end, zheng_meta-optic_2022, lin_all-optical_2018, wu2019phasecam3d, wei_spatially_2023}. In practice, this end-to-end (E2E) scheme requires excessive computational resources as the optical simulators are expensive to evaluate and the search space has a much higher dimension than purely optimizing computational models~\cite{wei_spatially_2023}. 

In this work, we propose an alternative strategy for designing opto-electronic \textit{convolutional} neural networks (CNNs) that alleviates the challenges of end-to-end training. Instead of optimizing the hybrid system jointly, we first train a conventional electronic CNN (or employ a pre-trained model) and then design the optical front-end implemented as a metasurface array to replicate its first convolutional layer through \textit{direct kernel optimization} (DKO). Compared to end-to-end optimization, this two-stage approach substantially simplifies the design process: the dimension of the variables to be optimized simultaneously is greatly reduced.

\begin{figure}
    \centering
    \includegraphics[width=\linewidth]{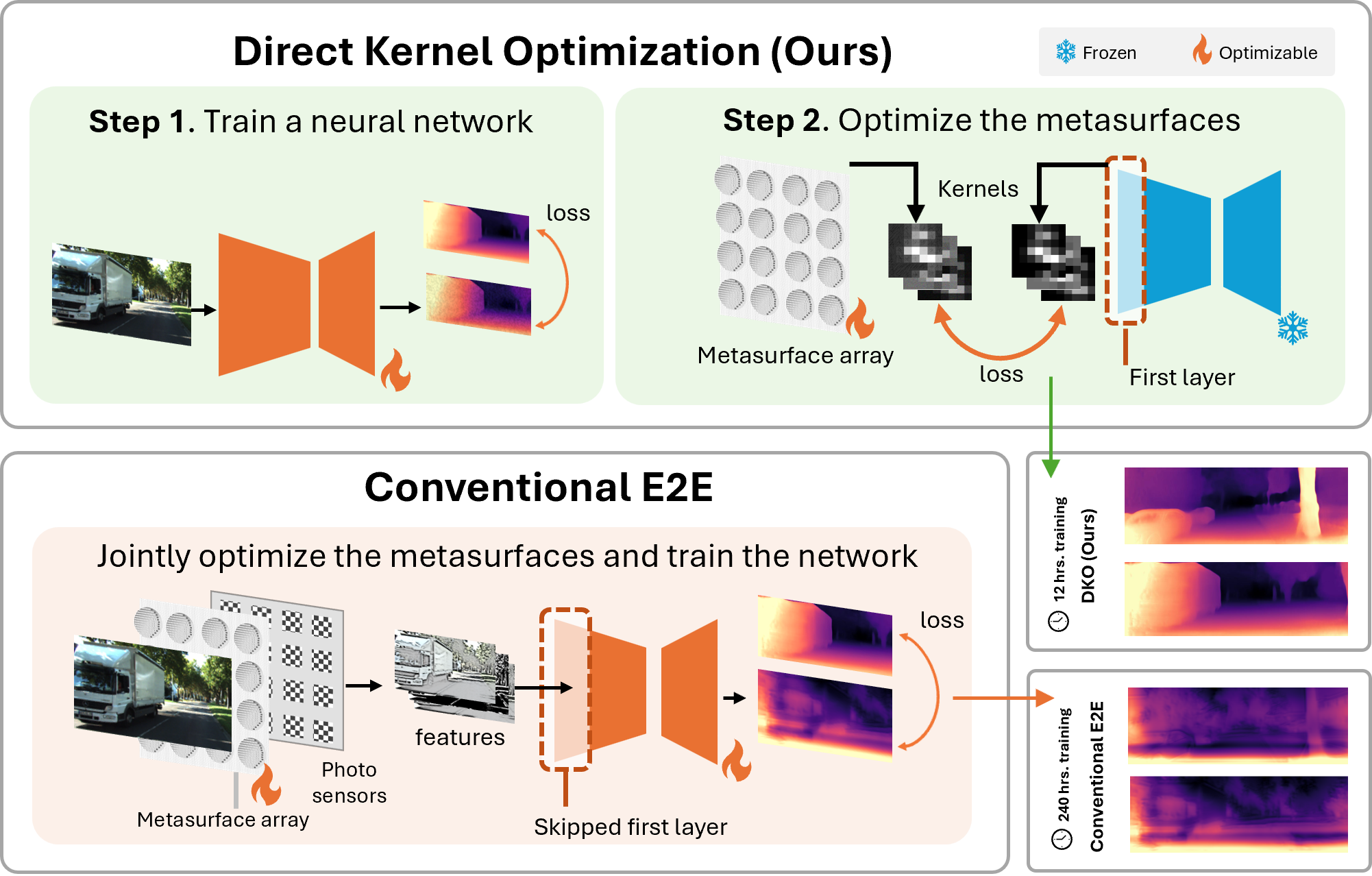}

    \caption{We consider an opto-electronic convolutional neural network (CNN) that integrates a metasurface array with an electronic backend. The metasurface encodes the incident light from a scene into optical feature maps. As light propagates through the metasurface, it undergoes a phase modulation equivalent to convolving a photograph of the scene with an engineered kernel. These optically generated feature maps are then processed electronically by a conventional CNN architecture. Our method (\textbf{top}) breaks down the optimization of these opto-electronic CNNs into tractable sub-problems, achieving better results in much less training time than conventional end-to-end (E2E) optimization of both the optics and the computational architecture (\textbf{bottom}).}
    \label{fig:overview}
\end{figure}

We demonstrate these advantages by designing and simulating opto-electronic CNNs for an exemplar task: monocular depth estimation. According to our analysis, the proposed two-stage method achieves $70\%$ higher accuracy than the joint optimization of optics and computation under uniform computational resource constraints. Furthermore, the dimension of the parameter space and computational cost of end-to-end training are hundreds of times higher, whereas the proposed approach maintains a significantly smaller computational footprint. 

Its key contributions include:
\begin{itemize}
\item A two-stage strategy to design opto-electronic CNNs for vision and imaging;
\item An exemplar opto-electronic CNN designed using the two-stage strategy for monocular depth estimation;
\item A comprehensive simulation study that demonstrates the accuracy, efficiency, and stability benefits of the proposed two-stage strategy over traditional end-to-end optimization.
\end{itemize}

\section{Related Work}

Incorporating optics into artificial vision and imaging systems has emerged as a vibrant discipline, fueled by recent advances in optical fabrication technologies that now enable the accessibility of custom devices such as diffractive optical elements~\cite{peng2015computational} and metasurfaces~\cite{chen2020flat}. Collectively referred to as computational optics, these devices form feature maps—rather than conventional photographs—on the photosensor. Such feature maps can be understood as scene embeddings, generated according to engineered sensitivities of the optical devices~\cite{wetzstein2020inference, fu2024optical}.

Based on their functionalities, these systems can be broadly divided into two categories. The first category exploits the optics’ intrinsic sensitivity to scene properties—such as depth~\cite{wu2019phasecam3d, chang2019deep}, spectrum~\cite{zheng_meta-optic_2022}, and polarization~\cite{hazineh2023polarization, rubin2019matrix}—to encode this information into the feature map through point spread functions (PSFs) that vary with the underlying scene attributes. The second category seeks to emulate part~\cite{wei_spatially_2023, zheng_meta-optic_2022, shi_loen_2022} or all~\cite{lin_all-optical_2018, xue2024fully} of a deep neural network architecture in the optical domain. Platforms in this class harness optics’ inherent speed and parallelism, employing one or more layers of optical arrays in which each element performs a linear transformation, such as a convolution, on the output of the preceding layer. Such fully or partially optical neural networks have been experimentally demonstrated on basic vision tasks, including image classification~\cite{zheng_meta-optic_2022, chang2018hybrid, bernstein2023single}.

These computational optics systems are often designed in an end-to-end (E2E) manner, where the optical elements and computational parameters are jointly optimized under a unified loss. Such co-optimization has been shown to yield superior local optima compared to separately designing the optics and the computation~\cite{wu2019phasecam3d, sitzmann2018end}. Nonetheless, implementing end-to-end optimization is challenging: the optical module requires differentiable solvers for light propagation—whether wave-based~\cite{hazineh_d-flat_2022, ho2024differentiable, tseng2021differentiable}, ray-based~\cite{wang2022differentiable, sun2021end}, or hybrid approaches~\cite{ren2025successive}—all of which are computationally intensive and significantly enlarge the design search space.

\section{System Design}

The proposed opto-electronic CNN, illustrated in Fig.~\ref{fig:overview}, employs a 2D metasurface array to simultaneously encode the scene into $M \times N$-channel optical feature maps on a shared photosensor. This metasurface layer functions as an optical approximation of the first convolutional layer of a pre-trained CNN. The resulting features are then fed into the subsequent electronic layers of the CNN, which process the features to generate the final output.

\subsection{Optical Model}

Consider an incoherent scene located at a distance much larger than the spatial extent of the metasurface array. 
The incident environmental light can be modeled as a superposition of incoherent plane waves with amplitude distribution $J(\vk)$ as a function of the wave vector $\vk=[k_x, k_y, k_z]$. 
Each planar wavefront right before the metasurface is expressed as:  
\begin{align}
    U(x,y;\vk) \approx A_0(\vk)\exp\left[j\left(k_xx + k_yy\right)\right],
\end{align}  
where $(x,y)$ denotes the coordinates on the metasurface array, and $A_0(\vk)$ is the amplitude of the plane wave.  

Each metasurface element $(m,n)$ is characterized by a modulation profile $C_{m,n}(x,y)$, which can be written as  
\begin{align}
    C_{m,n}(x,y) = T_{m,n}(x,y)\exp\left[j\varphi_{m,n}(x,y)\right],
\end{align}  
with $T_{m,n}(x,y)$ and $\varphi_{m,n}(x,y)$ representing the amplitude and phase modulation, respectively.  

The resulting power distribution generated by metasurface $(m,n)$ under an incident plane wave $\vk$ is determined by free-space propagation of the modulated wavefront~\cite{goodman2005introduction}:  
\begin{align}
    P_{m,n}(u,v;\vk) \propto A_0^2(\vk)
    \Bigg|\widetilde{C}_{m,n}\left(\frac{u-k_xs}{\lambda s},\frac{v-k_ys}{\lambda s}\right)\Bigg|^2,
    \label{eq:fresnel}
\end{align}  
where $\widetilde{C}_{m,n}$ denotes the Fresnel diffraction pattern of the modulated wavefront produced when a front-parallel plane wave propagates through the metasurface $C_{m,n}$~\cite{hazineh_d-flat_2022}. Eq.~\ref{eq:fresnel} indicates that the measurement formed on the photosensor, $I_{m,n}$, is a convolution of the pinhole image of the scene with an engineered kernel determined by the metasurface:
\begin{equation}
    \begin{aligned}
    I_{m,n}(u,v) &= \int_\vk P_{m,n}(u,v;\vk) d\vk \\
    &= I(u,v) * h_{m,n}(u,v), \\
    \text{where } I(u,v) &= \int_\vk A_0^2(\vk)d\vk \quad \text{ (pinhole image)}, \\
    h_{m,n}(u,v) &= \Bigg|\widetilde{C}_{m,n}\left(\frac{u}{\lambda s},\frac{v}{\lambda s}\right)\Bigg|^2 \quad \text{(kernel)}.  
\end{aligned}
\end{equation}
This property enables using metasurfaces to perform convolutional operations with the desired kernel $h_{m,n}$ by designing the modulation profiles $C_{m,n}$~\cite{zheng_meta-optic_2022, hazineh2023polarization}.

For convolutional kernels $h_{m,n}$ that contain negative values, we design two metasurfaces with modulation profiles $C_{m,n,+}$ and $C_{m,n,-}$, and approximate the kernel response by subtracting the two corresponding measurements. In addition, because metasurfaces are generally dispersive, the effective kernels vary with wavelength and are only partially correlated across the spectrum. To simplify the analysis, we restrict each metasurface to operate at a single wavelength of incident light. This can be practically achieved by placing a narrow bandpass filter in front of each metasurface.  

To extend the design to CNNs that process RGB images, where each kernel $h_{m,n}$ consists of three channels, we construct three independent pairs of metasurfaces $C_{m,n,\pm,R}$, $C_{m,n,\pm,G}$, and $C_{m,n,\pm,B}$. Each pair transmits only a narrow spectral band (red, green, or blue) from the scene and is assumed to implement a kernel that remains constant within that band. Consequently, to approximate the first convolutional layer with $L$ output channels for RGB inputs, the metasurface array requires $6L$ elements. 

The measurements formed by the metasurface array can be measured either using a photosensor array placed behind each metasurface (as shown in Fig.~\ref{fig:overview}) or a single shared photosensor.

\subsection{Direct Kernel Optimization}

We optimize a pair of metasurface phase modulation profiles, $\varphi_{m,n,+}(x,y)$ and $\varphi_{m,n,-}(x,y)$, assuming uniform transmittance profiles $T_{m,n,\pm}(x,y)$ within a predefined circular aperture, to approximate a given single-channel target kernel $h_{m,n}$. The optimization is formulated as  
\begin{equation}
    \underset{\varphi_{m,n,\pm}(x,y)}{\arg\min}
    \lVert \text{Simulator}(\varphi_{m,n,+}(x,y))
    - h_{m,n,\pm}(u,v) \rVert^2,
    \label{eq:okm}
\end{equation}
where  
\begin{equation*}
    h_{m,n,\pm}(u,v)=\frac{\pm h_{m,n}(u,v)+|\pm h_{m,n}(u,v)|}{2}.
\end{equation*}
We adopted the D-Flat differentiable simulator to generate the kernel given the phase modulation profiles~\cite{hazineh_d-flat_2022}. 
After determining the optimal phase modulation, $\varphi_{m,n,\pm}(x,y)$, we translate them into metasurface geometries by performing a standard cell-based library search.

\label{sec:typestyle}

\section{Results and Analysis}

In this paper, we focus on analyzing the proposed two-step strategy for designing opto-electronic CNNs and compare it with the traditional end-to-end strategy in simulation. 
The accuracy of the employed simulation process has been validated in our prior work~\cite{hazineh_d-flat_2022}, 
which showed that the simulated kernels closely matched those measured from fabricated metasurfaces 
designed with the same framework~\cite{hazineh2023polarization}.

We select monocular depth estimation as the target application for our study, and design the opto-electronic CNN based on a pre-defined architecture, Monodepth2~\cite{monodepth2}. 
This architecture takes a single RGB image as input, and its first convolutional layer contains 64 channels. 
Consequently, a total of $384=64\times 6$ metasurfaces need to be optimized to carry out the first layer operation optically.
For each metasurface $(m,n)$, the phase modulation profile $\varphi_{m,n}$ is parameterized as a $1025\times 1025$ discrete 2D matrix with a pixel pitch of $2.5\,\um \times 2.5\,\um$. 
The spacing between the metasurface array and the photosensor, i.e., the sensor distance, is set to $10$\,mm. 
The resolution of the feature maps generated by each metasurface is $320\times 96$. 
We adopt the KITTI dataset~\cite{geiger2013vision} for training, evaluation, and testing.

\begin{table}[]
    \centering
    \begin{tabular}{p{3.5cm} r r}  
    \hline
    \textbf{Method} & \textbf{Parameters (M)} & \textbf{Time (ms)} \\
    \hline
    DKO & 1.1 & 100 \\
    Computational Training & 14.84 & 250 \\
    E2E & 418 & 73,000 \\
    \hline
    \end{tabular}
    \caption{The number of trainable parameters (in millions), and the computational time (in milliseconds) for one forward pass and backward pass for our DKO method (first row), the computational training of Monodepth2 (second row), and the E2E method (third row).}
    \label{tab:computation}
\end{table}

\begin{figure}
    \centering
    \includegraphics[width=\linewidth]{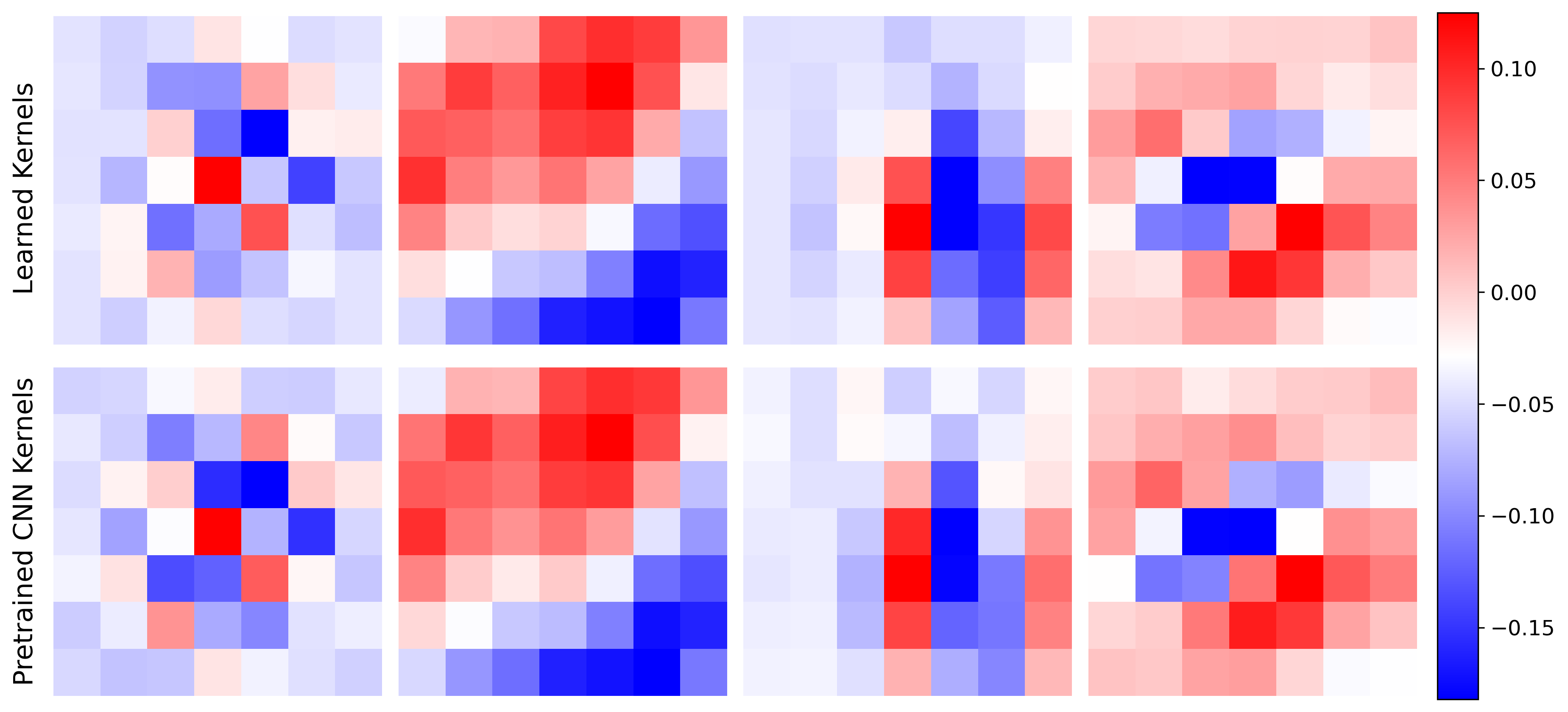}
    \caption{\textbf{Top row:} Sample metasurface-learned kernels $h_{m,n}(u,v)$, and \textbf{bottom row:}  corresponding kernels from the pretrained Monodepth2 model. Our optimized metasurfaces learn PSFs that closely match the original model's kernels.}
    \label{fig:predicted_kernels}
\end{figure}

\begin{figure*}
    \centering
    \setlength{\tabcolsep}{2pt}
    \begin{tabular}{@{\hskip 0.02\textwidth}ccccc@{\hskip 0.02\textwidth}}
        
        \includegraphics[width=0.18\textwidth]{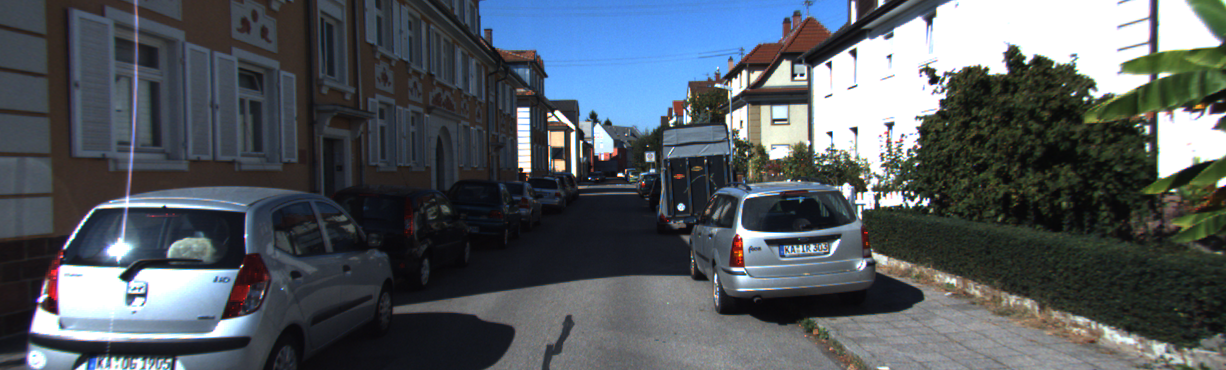} &
        \includegraphics[width=0.18\textwidth]{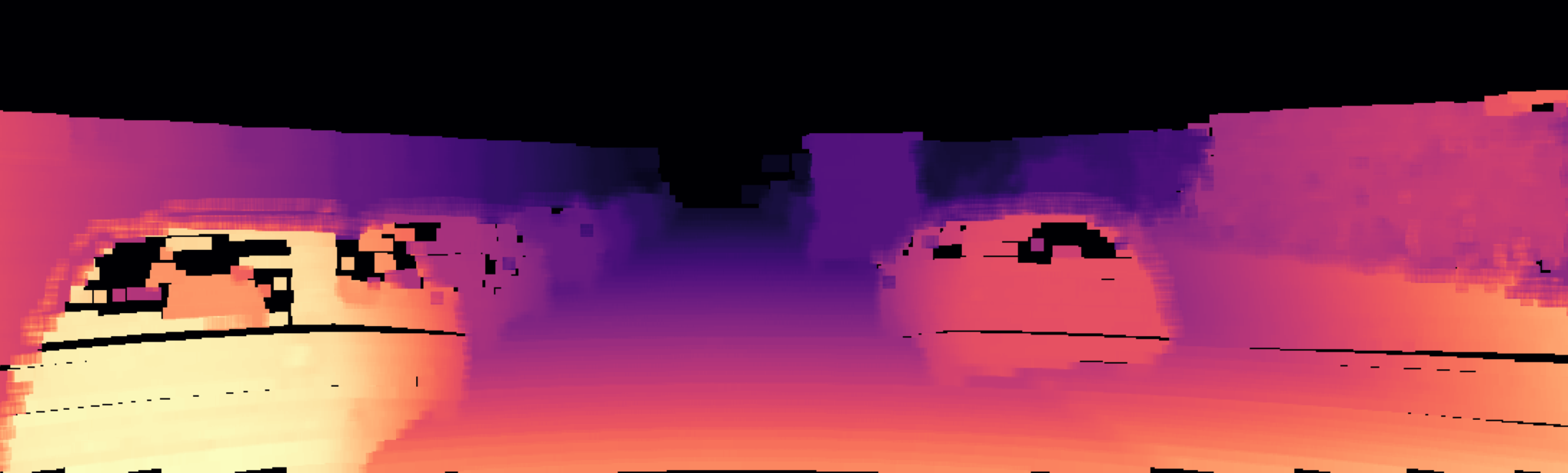} &
        \begin{overpic}[width=0.18\textwidth]{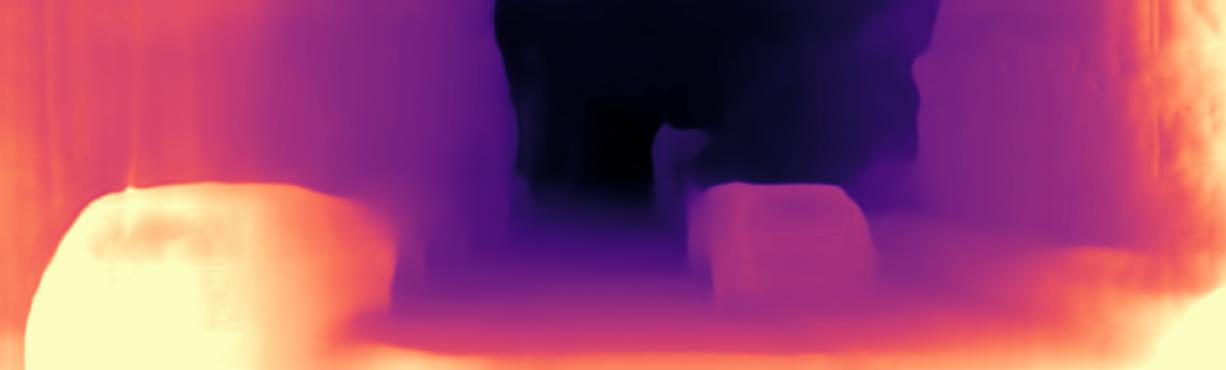}
            \put(63.1,23.0){\colorbox{black}{\color{white}\scriptsize \tiny RMSE: 4.818}}
        \end{overpic} &
        \begin{overpic}[width=0.18\textwidth]{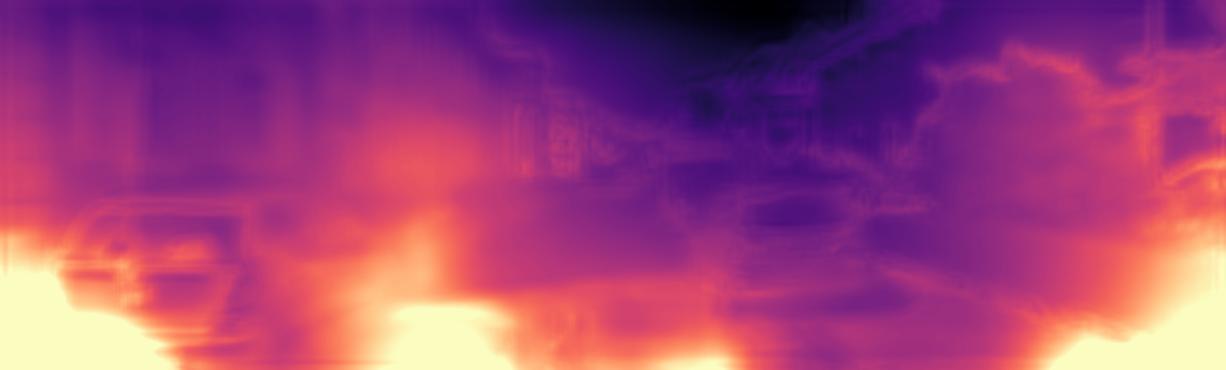}
            \put(63.1,23.0){\colorbox{black}{\color{white}\scriptsize \tiny RMSE: 9.357}}
        \end{overpic} &
        \begin{overpic}[width=0.18\textwidth]{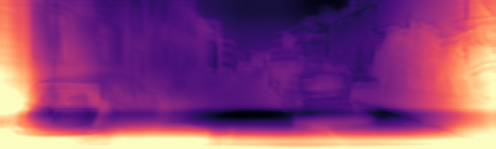}
            \put(63.1,23.0){\colorbox{black}{\color{white}\scriptsize \tiny RMSE: 9.594}}
        \end{overpic} \\

x           \includegraphics[width=0.18\textwidth]{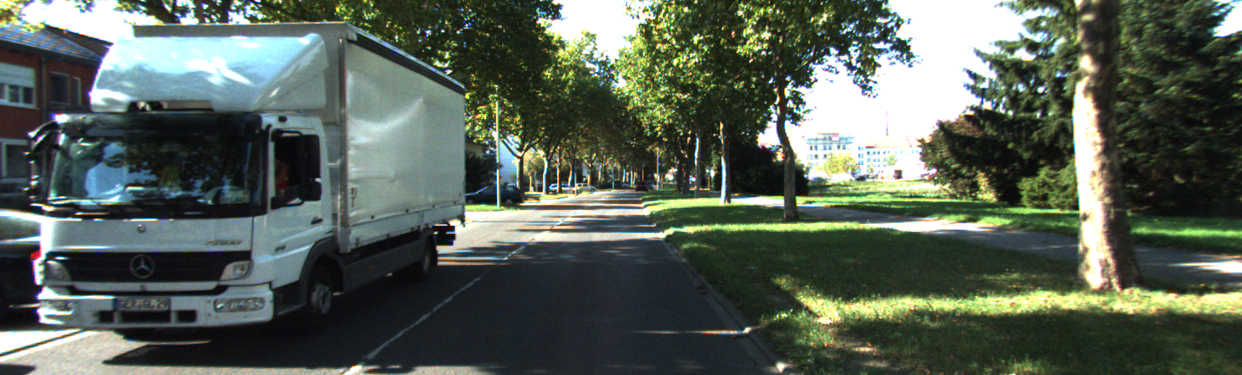} &
        \includegraphics[width=0.18\textwidth]{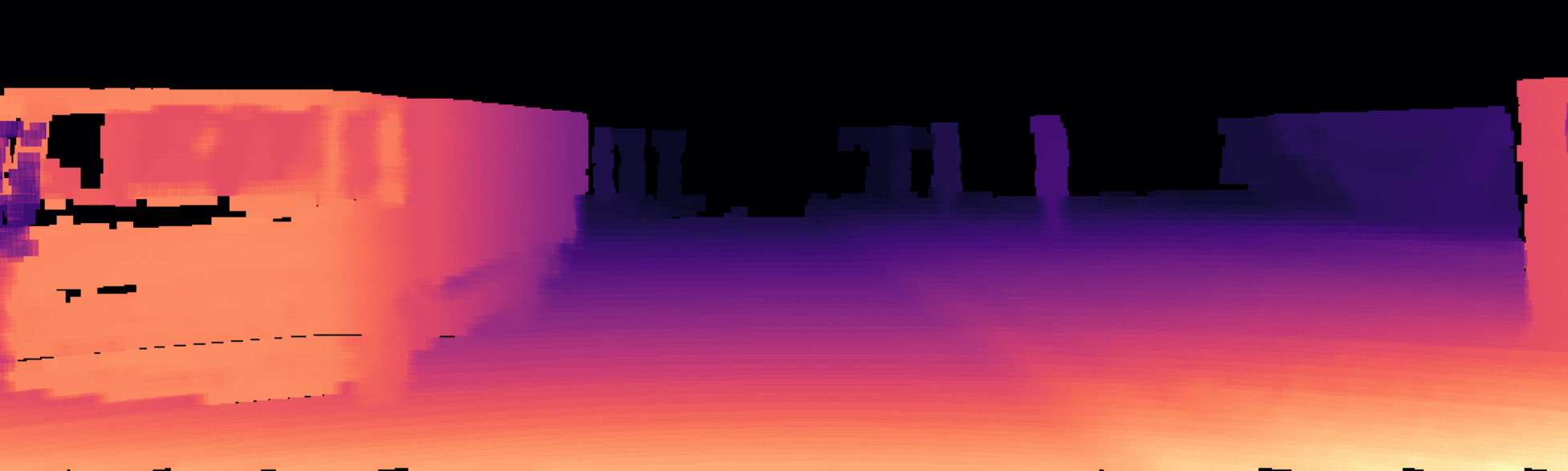} &
        \begin{overpic}[width=0.18\textwidth]{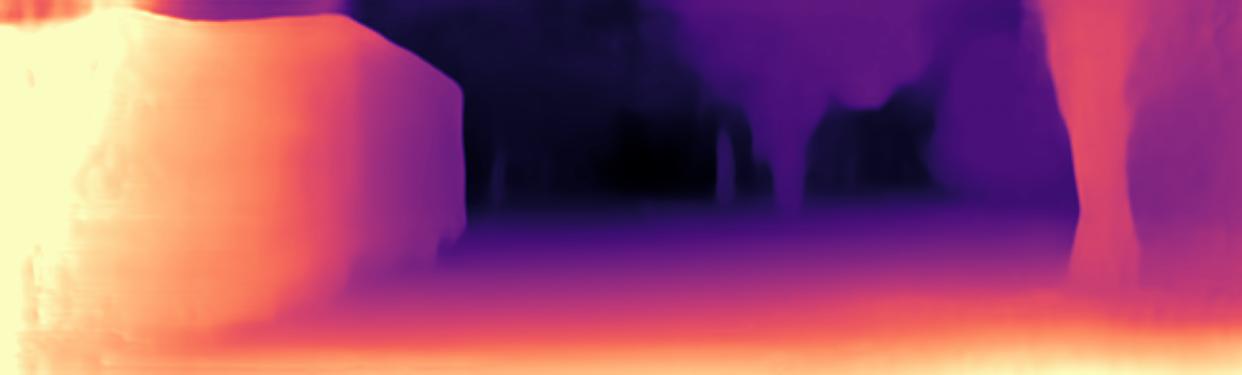}
            \put(63.1,23.1){\colorbox{black}{\color{white}\scriptsize \tiny RMSE: 5.735}}
        \end{overpic} &
        \begin{overpic}[width=0.18\textwidth]{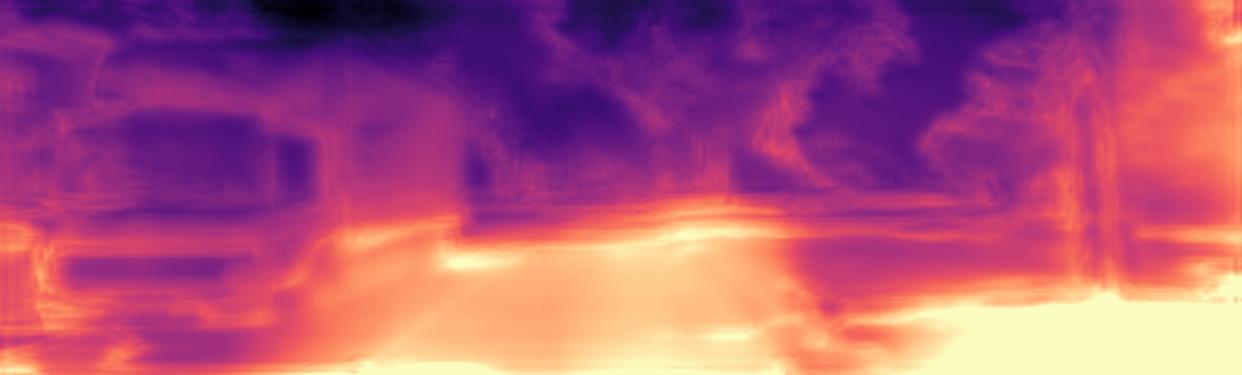}
            \put(60.5,23){\colorbox{black}{\color{white}\scriptsize \tiny RMSE: 12.130}}
        \end{overpic} &
        \begin{overpic}[width=0.18\textwidth]{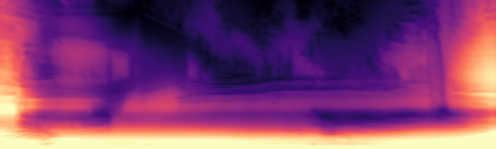}
            \put(60.5,22.9){\colorbox{black}{\color{white}\scriptsize \tiny RMSE: 12.194}}
        \end{overpic} \\

        \includegraphics[width=0.18\textwidth]{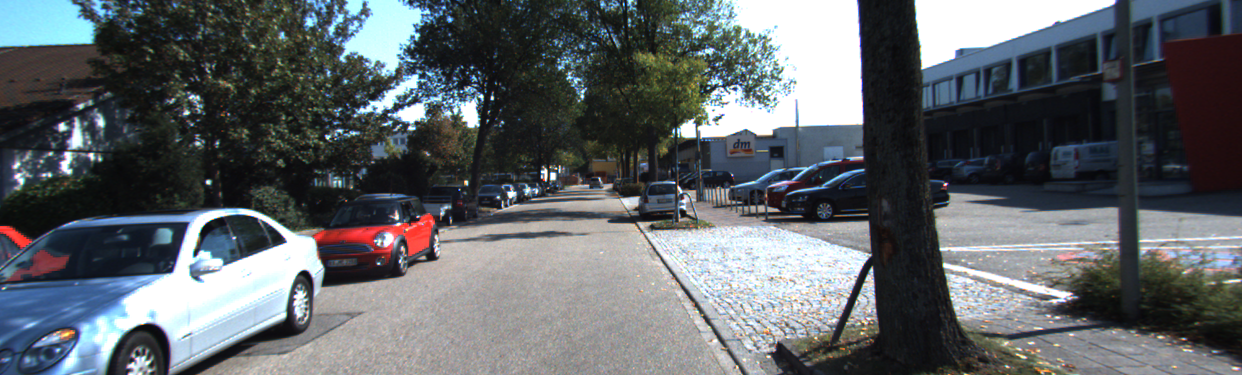} &
        \includegraphics[width=0.18\textwidth]{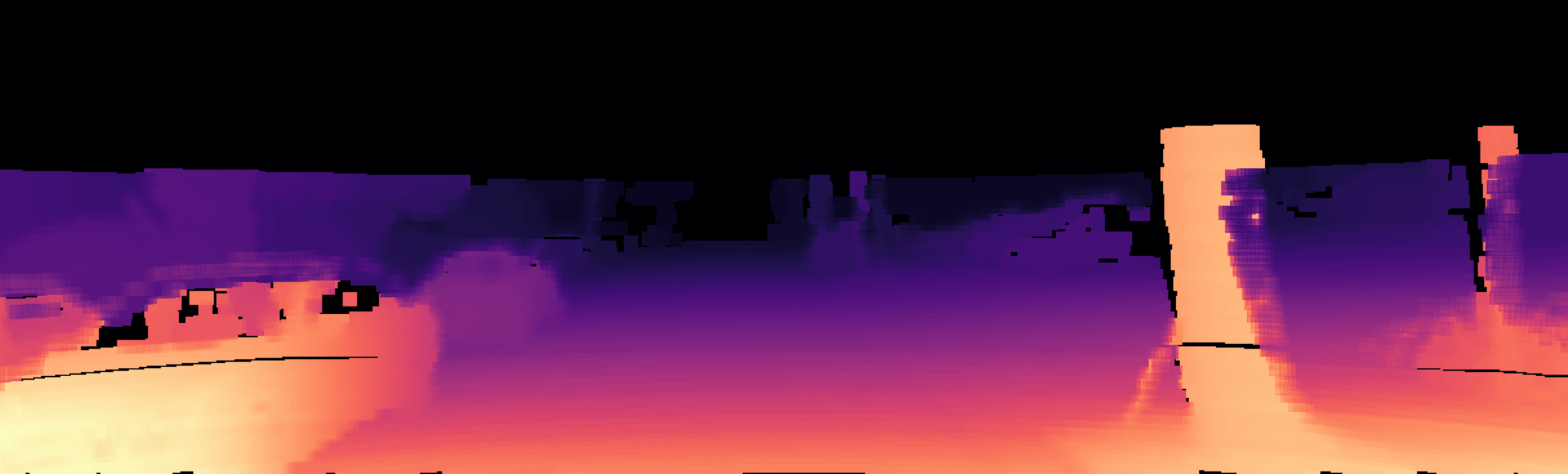} &
        \begin{overpic}[width=0.18\textwidth]{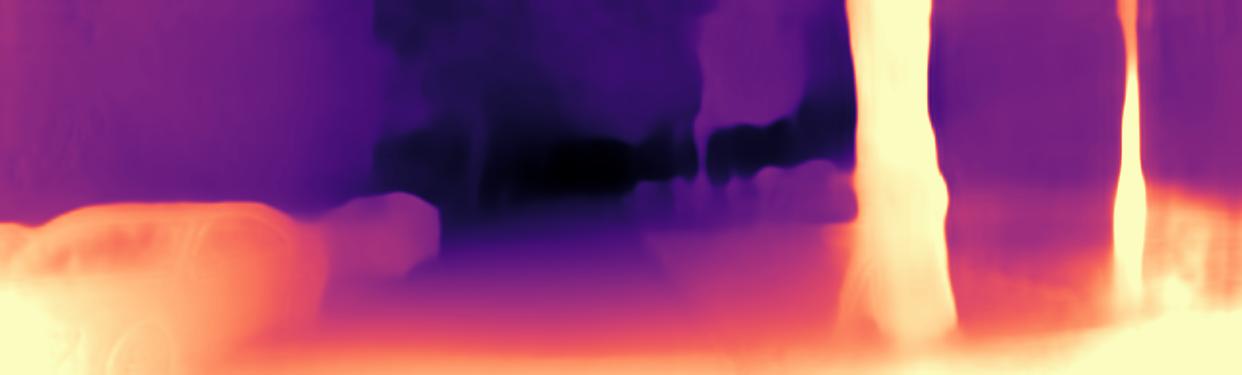}
            \put(63,23.1){\colorbox{black}{\color{white}\scriptsize \tiny RMSE: 7.978}}
        \end{overpic} &
        \begin{overpic}[width=0.18\textwidth]{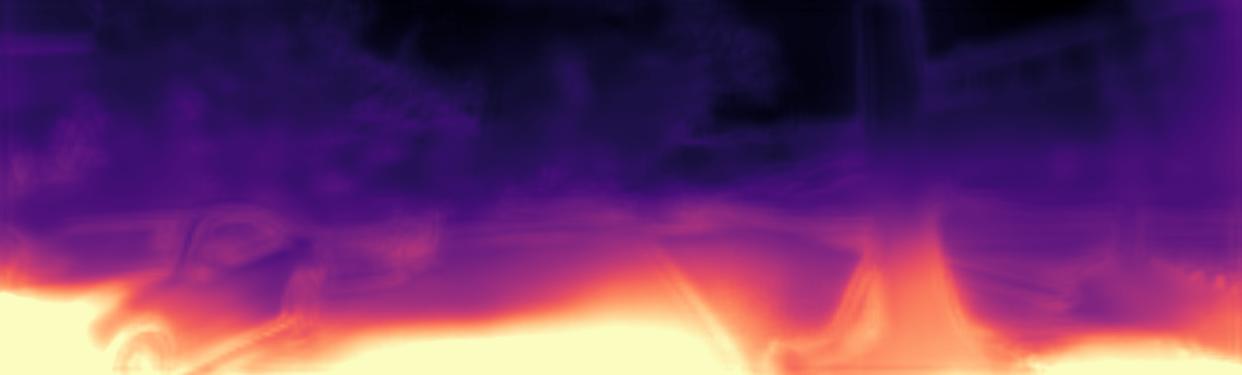}
            \put(60.5,23){\colorbox{black}{\color{white}\scriptsize \tiny RMSE: 12.549}}
        \end{overpic} &
        \begin{overpic}[width=0.18\textwidth]{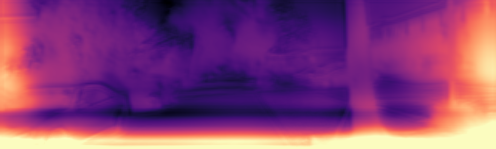}
            \put(60.5,23){\colorbox{black}{\color{white}\scriptsize \tiny RMSE: 13.859}}
        \end{overpic} \\

        \multicolumn{1}{c}{\textbf{Input}} &
        \multicolumn{1}{c}{\textbf{Ground Truth Depth}} &
        \multicolumn{1}{c}{\textbf{Our Method}} &
        \multicolumn{1}{c}{\textbf{E2E$_{\text{pretrained}}$}} &
        \multicolumn{1}{c}{\textbf{E2E$_{\text{w/o pretraining}}$}} \\

    \end{tabular}
    \caption{
    Qualitative comparison on the KITTI dataset (simulation). The first column shows the input image; the second column shows the sparse ground-truth depth map; the third column shows the result from a simulated opto-electronic CNN, where the first convolutional layer is implemented by a metasurface and trained using the proposed two-stage strategy; the fourth and fifth columns show results from the same system trained end-to-end, initialized with and without the pretrained model, respectively. All design strategies utilize uniform computational resources (one A100 GPU). The proposed two-stage strategy shows significantly better visual quality and accuracy compared to E2E strategies. 
    The inset numbers indicate the RMSE (in meters) for each prediction.
    }
    \label{fig:main_results}
\end{figure*}

\begin{table*}[]
     \centering
    \begin{tabular}{l c c c c c c c c}
        \hline
        \textbf{Experiment} & \textbf{AbsRel} & \textbf{SqRel} & \textbf{RMSE~(meters)} & \textbf{RMS$_{log}$} & $\boldsymbol{\delta < 1.25}$ & $\boldsymbol{\delta < 1.25^2}$ & $\boldsymbol{\delta < 1.25^3}$ & \textbf{Training time} \\
        \hline
        Ours            & \textbf{0.199} & \textbf{1.674} & \textbf{6.996} & \textbf{0.305} & \textbf{0.688} & \textbf{0.879} & \textbf{0.944} & \textbf{12h}    \\
        E2E$_{\text{pretrained}}$             & 0.346  &   3.618  &  11.013  &   0.494  &   0.401  &   0.675  &   0.835 & 240h \\
        E2E$_{\text{w/o pretraining}}$             &   0.387  &   4.173  &  11.853  &   0.552  &   0.324  &   0.633  &   0.814 & 240h \\
        
        \hline
    \end{tabular}
    \caption{Quantitative comparison on the KITTI dataset (simulation). Our two-stage strategy clearly achieves better performance than the E2E strategies, as the latter suffer from convergence issue due to the extremely high dimension of trainable variables. }
    \label{tab:depth_metrics}
\end{table*}

\paragraph*{Computational cost of training.} Under our parameter settings, the optical layer contains $403$M trainable parameters. 
By comparison, the first convolutional layer implemented electronically would require only $9$k parameters, 
and the entire Monodepth2 architecture has just $14$M parameters in total. 
Thus, an end-to-end training strategy would necessitate optimizing more than $400$M parameters jointly. 
In contrast, the proposed two-step strategy decouples the optimization of the optical layer from that of the electronic layers, 
and further breaks down the optical optimization into independent metasurface-level subproblems, 
substantially reducing the dimensionality of the search space (Table~\ref{tab:computation}.)

Moreover, end-to-end optimization requires rendering the full feature map of the scene using variants of Eq.~\ref{eq:fresnel}, 
whereas our two-step approach only evaluates the kernels (Eq.~\ref{eq:okm}) without the rendering step. 
This greatly reduces the computational burden during backpropagation. 
The computational advantages of the proposed two-step strategy are shown in Table~\ref{tab:computation}.

\paragraph*{Direct kernel optimization.} Our DKO approach effectively generates metasurfaces whose kernels closely match the target ones. As shown in Fig.~\ref{fig:predicted_kernels}, the learned kernels align well with the Monodepth2 target kernels, indicating that our metasurfaces would produce feature maps highly consistent with those of the original model. Note that the visualized kernels represent the final form, with negative components already subtracted from the positive ones. Table~\ref{tab:kernel_metrics} reports quantitative evaluations across all learned kernels, further validating this close correspondence.

\begin{table}[h]
    \centering
        \begin{tabular}{c c c}
        \hline
        NCC$\uparrow$ & RMSE $\downarrow$ & MAE $\downarrow$ \\
        \hline
        0.9840 & 0.012909 & 0.007555 \\
        \hline
        \end{tabular}
    \caption{Average normalized cross correlation (NCC), root mean square error (RMSE), and mean absolute error (MAE) for all $64\times6$ kernels produced by our optimized metasurface against the ground truth Monodepth2 first layer kernels.}
    \label{tab:kernel_metrics}
\end{table}
\vspace*{-1.5em}
\paragraph*{Depth estimation.} We trained the Monodepth2 architecture from scratch and optimized the metasurface kernels to match those in its first convolutional layer. The entire process completed in under 12 hours on a single Nvidia A100 GPU. In contrast, the E2E approach failed to produce meaningful results within the computational budget, owing to its prohibitive resource demands—even when the computational module was initialized to the pretrained model. Qualitative and quantitative comparisons between our strategy and the E2E baseline are shown in Fig.~\ref{fig:main_results} and Table~\ref{tab:depth_metrics}, respectively. The E2E optimization must jointly tune hundreds of millions of optical parameters, requiring repeated convolutions and backpropagation through the optical module for every batch.

\section{Conclusion}
We propose DKO, a two-stage framework that avoids the high cost of E2E training in hybrid opto-electronic CNNs. First, a purely computational CNN is trained conventionally; then a metasurface is directly optimized to reproduce the kernels of the first convolutional layer, effectively replacing that layer with an optical counterpart. Unlike prior hybrid approaches largely demonstrated on low-dimensional outputs (e.g., digit or object classification), our simulations show DKO supports dense prediction tasks such as monocular depth estimation, which require spatially resolved outputs. By decoupling optical design from full-network backpropagation, DKO substantially reduces training time and computational cost while maintaining the accuracy of the original computational model. The same framework naturally extends to other dense tasks, including semantic segmentation and surface-normal estimation, offering a practical path toward scalable hybrid vision systems that better leverage optical computing. 

\bibliographystyle{IEEEbib}
\footnotesize
\bibliography{strings,refs}

\end{document}

%% file: commands.tex
\providecommand{\vk}{\mathbf{k}}

\newcommand{\um}{\mu\text{m}}